\documentclass[%
 reprint,
 amsmath,amssymb,
 aps,
pra,
]{revtex4-1}

\usepackage{graphicx}
\usepackage{subfigure}
\usepackage{overpic}
\usepackage{dcolumn}
\usepackage{bm}
\usepackage{color}
\usepackage{soul}

\begin{document}

\preprint{APS/123-QED}

\title{Interacting Bose gas across a narrow Feshbach resonance}

\author{Fan Yang}
\author{Ran Qi}%
 \email{qiran@ruc.edu.cn}
\affiliation{%
 Department of Physics, Renmin University of China, Beijing, 100872, P. R. China
 \\}%

\date{\today}

\begin{abstract}
We use a two-channel model to investigate an interacting Bose gas across a narrow Feshbach resonance within a field path integral approach. The ground state properties show strong deviation from that of a broad Feshbach resonance or a single channel interaction. The deviation can be interpreted by the strong energy dependence of two-body scattering length near a narrow Feshbach resonance. As the density increases, the chemical potential and energy per particle are found to saturate while the inverse compressibility and phonon velocity undergo a significant reduction. We also take Gaussian fluctuations into account and calculate the ground state energy correction as well as the quantum depletion.
\begin{description}
\item[Usage]
Secondary publications and information retrieval purposes.
\end{description}
\end{abstract}

\maketitle


\section{\label{sec:level1}Introduction}

The studies of weakly interacting dilute Bose gas have a very long history ever since the pioneer works in the 1950s. A cornerstone progress has been made by Lee, Huang, and Yang in their famous paper \cite{LHY}. In their work, a low-density expansion was obtained for the equation of state at zero temperature:
\begin{eqnarray}
  \frac{E}{N}=\frac{2\pi\hbar^2a_s}{m}n\left(1+\frac{128}{15\sqrt\pi}\sqrt{na_s^3}\right),\label{LHYwide}
\end{eqnarray}
where $E/N$ is the energy per particle of the ground state, $n$ is the total density of the Bose gas and $a_s$ is the s-wave scattering length \cite{LHY}. The first term in Eq. (\ref{LHYwide}) can be obtained by a simple mean-field calculation while the second term includes the contribution from the zero-point energy of quasi-particle excitations above the mean-field ground state which is often called Lee-Huang-Yang correction nowadays \cite{Pethick}. Based on Eq. (\ref{LHYwide}), all the thermal dynamic quantities can be derived at zero temperature. For example, the chemical potential and compressibility are given as:
 \begin{eqnarray}
  \mu=&&\left(\frac{\partial E}{\partial N}\right)_{V}=\frac{4\pi\hbar^2a_s}{m}n\left(1+\frac{32}{3\sqrt{\pi}}\sqrt{na_s^3}\right)\label{LHYwidemu},
  \\
  \kappa^{-1}=&&\left(\frac{\partial \mu}{\partial n}\right)_{N}=\frac{4\pi\hbar^2a_s}{m}\left(1+\frac{16}{\sqrt{\pi}}\sqrt{na_s^3}\right).\label{LHYwidekappa}
 \end{eqnarray}
According to Eq. (\ref{LHYwide})-(\ref{LHYwidekappa}), as the density increases with fixed $a_s$, both the energy per particle and chemical potential increase monotonically and faster than linear as a function of $n$, while the inverse compressibility remains a constant at the mean-field level and shows a weak dependence on density when including the Lee-Huang-Yang correction.
Although these results were published more than 60 years ago, their experimental verification is only made possible very recently, thanks to the realization and high precision measurement of Bose-Einstein condensate in ultracold quantum gases \cite{BEC,BEC2,LHYexp}.

However, one should note that the validity of expansion (\ref{LHYwide}) does not only require the low-density condition $n^{1/3}a_s\ll1$ but also relies on the fact that the two-body scattering process is fully determined by a single parameter $a_s$. For quantum gases close to a Feshbach resonance, while the second condition is usually fulfilled for a broad Feshbach resonance, it may break down for a very narrow resonance \cite{FiniteRange1,FiniteRange2,FiniteRangeGP1,FiniteRangeGP2,FiniteRangeGP3,FiniteRangeField1,FiniteRangeField2}.
As shown in several previous works, near a Feshbach resonance, the two-body scattering amplitude is generally determined by an energy-dependent scattering length $a_s(E)=a_{bg}[1+\alpha_r^2g_r^{-1}/(E-\nu_r)]$ where $E$ is the total energy of the two particles under collision in the center of mass frame \cite{FR,narrowFR,narrowFR1,narrowFR2}. Across a broad resonance, the energy dependence in $a_s(E)$ can be safely neglected and one has $a_s(E)\simeq a_s(0)$. In contrast, for a narrow resonance $a_s(E)$ has a very sensitive dependence on $E$ and thus the full functional form must be taken into account \cite{narrowFR,narrowFR1,narrowFR2,narrowFR3,narrowFR4}.

\begin{figure}[htbp]
\centering
\includegraphics[width=1\linewidth]{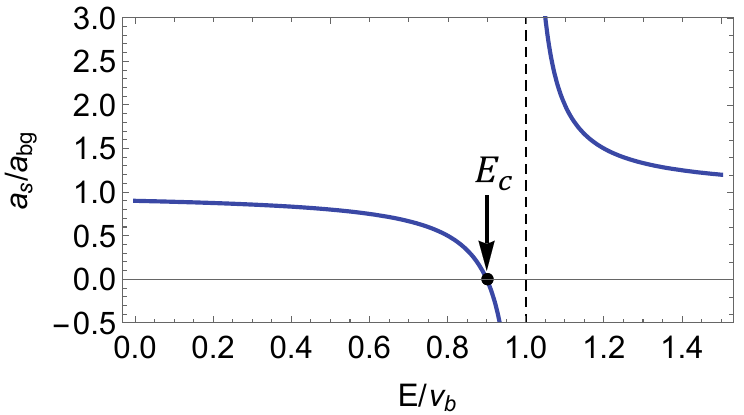}
\caption{Schematic plot of energy-dependent scattering length near a narrow Feshbach resonance.}\label{Fig1}
\end{figure}

In particular, for a Bose gas across a narrow Feshbach resonance, one can qualitatively estimate the effect of energy-dependent scattering length as follows. Since the chemical potential $\mu$ is the lowest energy to excite a single particle out of the condensate, the typical two-body scattering energy can be estimated as $E\sim2\mu$. As a result, the effective interacting strength should be determined by $a_s(2\mu)$ instead of the zero energy scattering length $a_s(0)$. In this paper, we consider a particular case $\nu_r>0, a_{bg}>0$, and a schematic plot of $a_s(E)$ is shown in Fig. \ref{Fig1}. In this case, the function $a_s(E)$ decreases from $a_s(0)>0$ to 0 as energy $E$ increases from 0 to a critical value $E_c$. At a fixed magnetic field, if one increases the density, the chemical potential $\mu$ will increase according to Eq. (\ref{LHYwidemu}) such that $a_s(2\mu)$ decreases which tends to prevent $\mu$ from further increasing. As a result, we expect that $\mu$ may saturate to $E_c/2$ as the density increases which leads to a nearly zero inverse compressibility. In the following, we will show that the qualitative analysis above is correct based on a mean-field plus Gaussian fluctuation calculation. We will provide a modified low-density expansion for the equation of state which can be seen as a generalization of Eq. (\ref{LHYwide}) that applies to Bose gases across either broad or narrow Feshbach resonances. The effect of energy-dependent scattering length on various thermodynamic properties will be investigated systematically.

Our paper is organized as follows. In Sec. \ref{sec:level2}, we introduce our model for a spinless Bose gas across a narrow Feshbach resonance and adopt the path integral approach to obtain the thermodynamic partition function \cite{pathIM} .  In Sec. \ref{sec:level3} we present the mean-field results on thermodynamic properties in detail and show the effect of energy-dependent scattering length in uniform systems. In Sec. \ref{sec:level4}, we calculate the Gaussian fluctuation correction to the mean-field results which does not show qualitative changes. In Sec. \ref{sec:level5}, we investigate density distributions for a trapped Bose gas. In Sec. \ref{sec:level6}, we summarize our main findings and conclude the paper.

\section{Model and path integral approach\label{sec:level2}}
For spinless bosons interacting across a magnetic Feshbach resonance, we adopt the widely used two-channel model:
 \begin{subequations}
  \begin{align}
   \hat H =& \hat H_a + \hat H_m + \hat H_c,\label{Htot}\\
   \hat H_a =&\int\mathrm d^3\mathbf x\left[\psi^\dagger(\mathbf x)\left(-\frac{\nabla^2}{2m}+V_a(\mathbf x)\right)\psi(\mathbf x)\right.\nonumber\\
   &+\left. \frac{g}{2}\psi^\dagger(\mathbf x)\psi^\dagger(\mathbf x)\psi(\mathbf x)\psi(\mathbf x)\right],\\
   \hat H_b =& \int\mathrm d^3\mathbf x\ b^\dagger(\mathbf x)\left(-\frac{\nabla^2}{2M}+\nu_b+V_b(\mathbf x)\right)b(\mathbf x),\\
   \hat H_c =& \int\mathrm d^3\mathbf x\left[\frac{\alpha}{\sqrt 2}\psi^\dagger(\mathbf x)\psi^\dagger(\mathbf x) b(\mathbf x)+h.c.\right],
  \end{align}
 \end{subequations}
where $\psi^\dagger$ and $b^\dagger$ are creation field operators for atoms and molecules respectively, and we set $\hbar=1$ throughout this paper. $V_a$ and $V_b$ are external potentials of atoms and molecules. The molecule detuning $\nu_b$, the inter-channel coupling $\alpha$, and the interaction parameter in open channel $g$ are bare quantities which need to be renormalized as follows:
 \begin{subequations}
   \begin{align}
    \nu_b\ &= \nu_{\mathrm r}-[1-Z(\Lambda)] \alpha_{\mathrm r}^2/g_{\mathrm r},\\
    \alpha\ &=\ Z(\Lambda) \alpha_{\mathrm r},\\
    g\ &=\ Z(\Lambda) g_{\mathrm r},
   \end{align}\label{renormalization}
 \end{subequations}
where $Z(\Lambda)=\left(1-g_{\mathrm r}\Lambda\right)^{-1}$ and $\Lambda=\frac{1}{V}\sum_{\mathbf k}'\left(2\varepsilon_{\mathbf k}\right)^{-1}$. The renormalized quantities $g_{\mathrm r}$ , $\alpha_{\mathrm r}$, and $\nu_{\mathrm r}$ determine the energy-dependent scattering length as
 \begin{equation}
   \frac{4 \pi a_s(E)}{m}=g_{\mathrm r}+\frac{\alpha_{\mathrm r}^2}{E-\nu_{\mathrm r}}.
 \end{equation}
The partition function of the Hamiltonian Eq. (\ref{Htot}) at arbitrary temperature $T$ can be written into the following imaginary time path integral form
 \begin{gather}
    Z\ =\ \int\mathrm D[\phi^*]\mathrm D[\phi]\mathrm D[\beta^*]\mathrm D[\beta]\exp({-S}),\label{Zpartition}
 \end{gather}
where the action $S$ is given as
  \begin{subequations}
    \begin{align}
      S =& \int\mathrm d x\left[\phi^*(x) \frac{\partial}{\partial\tau}\phi(x) + \beta^*(x) \frac{\partial}{\partial\tau}\beta(x)\right.\nonumber\\
      &+h_a(x) + h_b(x) + h_c(x)\bigg],\label{action}\\
      h_a(x) =& \phi^*(x)\left[-\frac{\nabla^2}{2m}+V_a(\mathbf x)-\mu\right]\phi(x),\nonumber\\
      &+ \frac{g}{2}\phi^*(x)\phi^*(x)\phi(x)\phi(x),\\
      h_b(x) =& \beta^*(x)\left[-\frac{\nabla^2}{4m}+ \nu_b +V_b(\mathbf x)-2\mu\right]\beta(x),\\
      h_c(x) =&  \frac{\alpha}{\sqrt 2}\phi^*(x)\phi^*(x)\beta(x)+\mathrm{c.c.},
   \end{align}
  \end{subequations}
and we have defined $x\equiv(\mathbf x,\tau)$ and $\int dx\equiv\int d^3\mathbf x\int_0^{\beta}d\tau$ with $\beta=(k_BT)^{-1}$.

The path integral (\ref{Zpartition}) can not be performed exactly due to the interaction terms. However, for a weakly interacting Bose gas with very low density, it is a good approximation to expand the action S in Eq. (\ref{action}) around its saddle point solution \cite{pathIM}:
  \begin{align}
    \phi(x) = \phi_0(\mathbf x)+\phi'(x),\\
    \beta(x) = \beta_0(\mathbf x)+\beta'(x) ,
  \end{align}
where $\phi_0(\mathbf x)$ and $\beta_0(\mathbf x)$ are the saddle point solution which minimizes the action $S$ while $\phi'(x)$ and $\beta'(x)$ are the fluctuation fields of atoms and molecules.

Later in Sec.\ref{sec:level3}, we will take the mean-field approximation by neglecting all the fluctuation terms. The approximate action is then given by
  \begin{align}
    S^{(0)}
      =& \beta\int\mathrm d^3\mathbf x\left\{\phi_0^*(\mathbf x)\left[-\frac{\nabla^2}{2m}+V_a(\mathbf x)-\mu\right]\phi_0(\mathbf x)\right.\nonumber\\
      &+ \frac{g}{2}|\phi_0(\mathbf x)|^4 + \beta_0^*(\mathbf x)\left[-\frac{\nabla^2}{4m}+V_b(\mathbf x)-2\mu\right]\beta_0(\mathbf x)\nonumber\\
      &\left. +\frac{\alpha}{\sqrt 2}\phi_0^*(\mathbf x)^2\beta_0(\mathbf x)+\frac{\alpha^*}{\sqrt 2}\beta_0^*(\mathbf x)(\phi_0(\mathbf x))^2\right\}.\label{action mf}
  \end{align}

Then in Sec.\ref{sec:level4} we will include the contribution from Gaussian fluctuation around the saddle point. For a uniform system, the saddle point solution is uniform, i.e. $\phi_0(\mathbf x)\equiv\phi_0$, $\beta_0(\mathbf x)\equiv\beta_0$, and we obtain the quadratic action as
  \begin{eqnarray}
    \quad S \approx S^{(0)}+S^{(2)},
  \end{eqnarray}
  \begin{align}
    S^{(2)}
      &= \int\mathrm dx \phi'^*(x)\left[ \frac{\partial}{\partial \tau}-\frac{\nabla^2}{2m}-\mu+2g|\phi_0|^2\right]\phi'(x)\nonumber\\
      &+ \frac{1}{2}\int\mathrm dx \ \left\{\phi'^*(x)^2\left[g\phi_0^2+\sqrt{2}\alpha\beta_0\right]+\mathrm{c.c.}\right\}\nonumber\\
      &+ \int\mathrm dx \beta'^*(x)\left[ \frac{\partial}{\partial\tau}-\frac{\nabla^2}{4m}-2\mu+\nu_b\right]\beta'(x)\nonumber\\
      &+ \int\mathrm dx \left[\
      2\frac{\alpha}{\sqrt 2}\phi'^*(x)\phi_0^*\beta'(x) + \mathrm{c.c.}\right].
  \end{align}
To perform the path integral for this quadratic action, it is more convenient to first transform the action into momentum-frequency space,
  \begin{align}
      \phi'(x) = \frac{1}{\sqrt{\beta V}}\sum_{\mathbf{k},n}\phi'_{\mathbf k,n}\exp[i(\mathbf k\cdot\mathbf x-\omega_n\tau)],\\
      \beta'(x) = \frac{1}{\sqrt{\beta V}}\sum_{\mathbf{k},n}\beta'_{\mathbf k,n}\exp[i(\mathbf k\cdot\mathbf x-\omega_n\tau)],
    \end{align}
where $\omega_n$ is the bosonic Matsubara frequency, and $\mathbf k$ is free wave vector. Then the action can be written more compactly as a matrix multiplication in Nambu space
  \begin{align}
    S^{(2)}
      =&-\frac{\beta}{2}\sum_{\mathbf k\neq 0}(\epsilon_{\mathbf k}^b+\epsilon_{\mathbf k}^a+2g|\phi_0|^2)\nonumber\\
      &-\frac{1}{2}\sum_{\mathbf k\neq0,n}\Phi_{k}^\dagger\mathbf G^{-1}(\mathbf k,i\omega_n)\Phi_{k},\label{action fluc}
  \end{align}
where $k\equiv \left(\mathbf k,i\omega_n\right)$, $\epsilon_{\mathbf k}^b=\frac{{\mathbf k}^2}{4m}+\nu_b-2\mu$, $\epsilon_{\mathbf k}^a=\frac{{\mathbf k}^2}{2m}-\mu$  and we defined the following vector
  \begin{equation}
    \Phi_{k}^\dagger = \big[\beta'^*_{k},\phi'^*_{k},\phi'_{-k},\beta'_{-k}\big].
  \end{equation}
The first summation in Eq. (\ref{action fluc}) comes from the order exchange between the creation and annihilation fields during the transformation into Nambu space \cite{pathIM}. And the $4\times 4$ matrix $\mathbf G^{-1}$ gives the inverse Green's function
  \begin{equation}
    \mathbf G^{-1} = \mathbf G_0^{-1} - \bm\Sigma,\label{inverseG}
  \end{equation}
where
  \begin{equation}
    \mathbf G_0^{-1} =
    \begin{bmatrix}
      G_{b0}^{-1}(k)&\ &\ &\ \\
      \ &G_{a0}^{-1}(k)&\ &\ \\
      \ &\ &G_{a0}^{-1}(-k)&\ \\
      \ &\ &\ &G_{b0}^{-1}(-k)
    \end{bmatrix},
  \end{equation}
  and
  \begin{equation}
    \bm\Sigma =
    \begin{bmatrix}
      0 &\tilde\alpha^* &\ &\ \\
      \tilde\alpha & 2g|\phi_0|^2 & \tilde g &\ \\
      \ &\tilde g* & 2g|\phi_0|^2 & \tilde\alpha^*\\
      \ &\ &\tilde\alpha & 0\\
    \end{bmatrix}.
  \end{equation}
Here we have defined $G_{a0}(k)=\left(i\omega_n-\epsilon_{\mathbf k}^a\right)^{-1}$, $G_{b0}(k)=\left(i\omega_n-\epsilon_{\mathbf k}^b\right)^{-1}$, $\tilde\alpha=\sqrt 2\alpha\phi_0^*$, and $\tilde g=g\phi_0^2+\sqrt 2\alpha\beta_0$.

It is known that the quasi-particle excitation corresponds to the poles of the Green's function. By diagonalizing Eq. (\ref{inverseG}) we obtain two branches of excitations given as
\begin{equation}
 \omega_{\mathbf k}^{\pm}= \sqrt{\left(B\pm\sqrt{B^2-4C}\right)/2},\label{quasiexcitation}
\end{equation}
where
  \begin{align}
      B\ &=\ \left(\epsilon_{\mathbf k}^a+2g|\phi_0|^2\right)^2 + \left(\epsilon_{\mathbf k}^b\right)^2 + 2|\tilde\alpha|^2-\mu^2,\\
      C\ &=\ \left[\left(\epsilon_{\mathbf k}^a+2g|\phi_0|^2\right)\epsilon_{\mathbf k}^b-|\tilde\alpha|^2\right]^2 - \mu^2\left(\epsilon_{\mathbf k}^b\right)^2.
    \end{align}
It is straightforward to check that $\omega_{\mathbf k}^{-}$ has a linear dependence on $|\mathbf{k}|$ as $\mathbf{k}\rightarrow 0$ and thus represents the phonon mode of this Bose superfluid. On the other hand, $\omega_{\mathbf k}^+$ is gapped at $\mathbf{k}=0$ corresponding to density fluctuation of closed channel molecule.

The Green's function can be diagonalized with a transformation matrix  $\mathbf U$ \cite{supple}
  \begin{equation}
    \begin{aligned}
      -&\mathbf U^{\mathbf T}\mathbf G^{-1}(\mathbf k,i\omega_n)\mathbf U =\\
      &\begin{bmatrix}
        -i\omega_n + \omega_{\mathbf k}^+&\ &\ &\ \\
        \ &-i\omega_n + \omega_{\mathbf k}^-&\ &\ \\
        \ &\ &i\omega_n + \omega_{\mathbf k}^-&\ \\
        \ &\ &\ &i\omega_n + \omega_{\mathbf k}^+\\
      \end{bmatrix}.\label{Gmatrix}
    \end{aligned}
  \end{equation}
Finally, we obtain the following Gaussian action
  \begin{align}
      S
      =& S^{(0)} +\frac{\beta}{2}\sideset{}{'}\sum_{\mathbf k}\left(\omega_{\mathbf k}^++\omega_{\mathbf k}^-\right)\nonumber\\
      &- \frac{\beta}{2}\sideset{}{'}\sum_{\mathbf k}\left(\epsilon_{\mathbf k}^a+2g|\phi_0|^2+\epsilon_{\mathbf k}^b\right)\nonumber\\
      &+ \sideset{}{'}\sum_{\mathbf k}(-i\omega_n + \omega_{\mathbf k}^+)\beta_{\mathbf k,n}^*\beta_{\mathbf k,n}\nonumber\\
      &+ \sideset{}{'}\sum_{\mathbf k}(-i\omega_n + \omega_{\mathbf k}^-)\phi_{\mathbf k,n}^*\phi_{\mathbf k,n},\label{action diagonalized}
    \end{align}
where $\big[\beta^*_{k},\phi^*_{k},\phi_{-k},\beta_{-k}\big]$ is related to $\big[\beta'^*_{k},\phi'^*_{k},\phi'_{-k},\beta'_{-k}\big]$ through
  \begin{equation}
    \big[\beta^*_{k},\phi^*_{k},\phi_{-k},\beta_{-k}\big]=\big[\beta'^*_{k},\phi'^*_{k},\phi'_{-k},\beta'_{-k}\big]\mathbf U^{-1}.
  \end{equation}
Again the summation of $\omega^+_{\mathbf k}$ and $\omega^-_{\mathbf k}$ in Eq. (\ref{action diagonalized}) comes from the exchange of the field operators \cite{pathIM}.

In the following sections we calculate the zero temperature thermodynamic potential $\Omega$ and the total density $n$ with the action given by either Eq. (\ref{action mf}) (in Sec. \ref{sec:level3}) or Eq. (\ref{action diagonalized}) (in Sec. \ref{sec:level4}) through following thermodynamic relations
\begin{gather}
    \frac{\Omega}{V}\ =\ -\frac{1}{\beta V}\ln(Z)\label{thermodynamic potential},\\
    n\ =\ -\frac{\partial (\Omega/V)}{\partial \mu},\label{densitytotal}\\
    \frac{E}{V}\ =\ \frac{\Omega}{V} + n\mu.
\end{gather}

\section{\label{sec:level3}Mean-field calculation}

In this section, we neglect the contribution of Gaussian fluctuation, and the action is governed by Eq. (\ref{action mf}). After minimizing the action, we obtain the two-channel G-P equations
\begin{subequations}
  \begin{align}
    &\left[\!-\!\frac{\nabla^2}{2m}\!+\!V_a\!+\!g_r|\phi_0(\mathbf x)|^2\!-\!\mu\!\right]\phi_0(\mathbf x)\!+\!\sqrt 2 \alpha_r\phi_0^*(\mathbf x)\beta_0(\mathbf x)\!=\!0\\
    &\left[-\frac{\nabla^2}{4m}+\nu_r+V_b-2\mu\right]\beta_0(\mathbf x)+\frac{\alpha_r^*}{\sqrt 2}\phi_0(\mathbf x)^2=0.
  \end{align}\label{GPequation}
\end{subequations}
The renormalization should be taken to the same order, so that at mean-field level we should use the renormalized parameters directly in Eq. (\ref{GPequation}) \cite{pathIM}.

\begin{figure}[htbp]
  \centering
  \begin{overpic}[width=1.\linewidth]{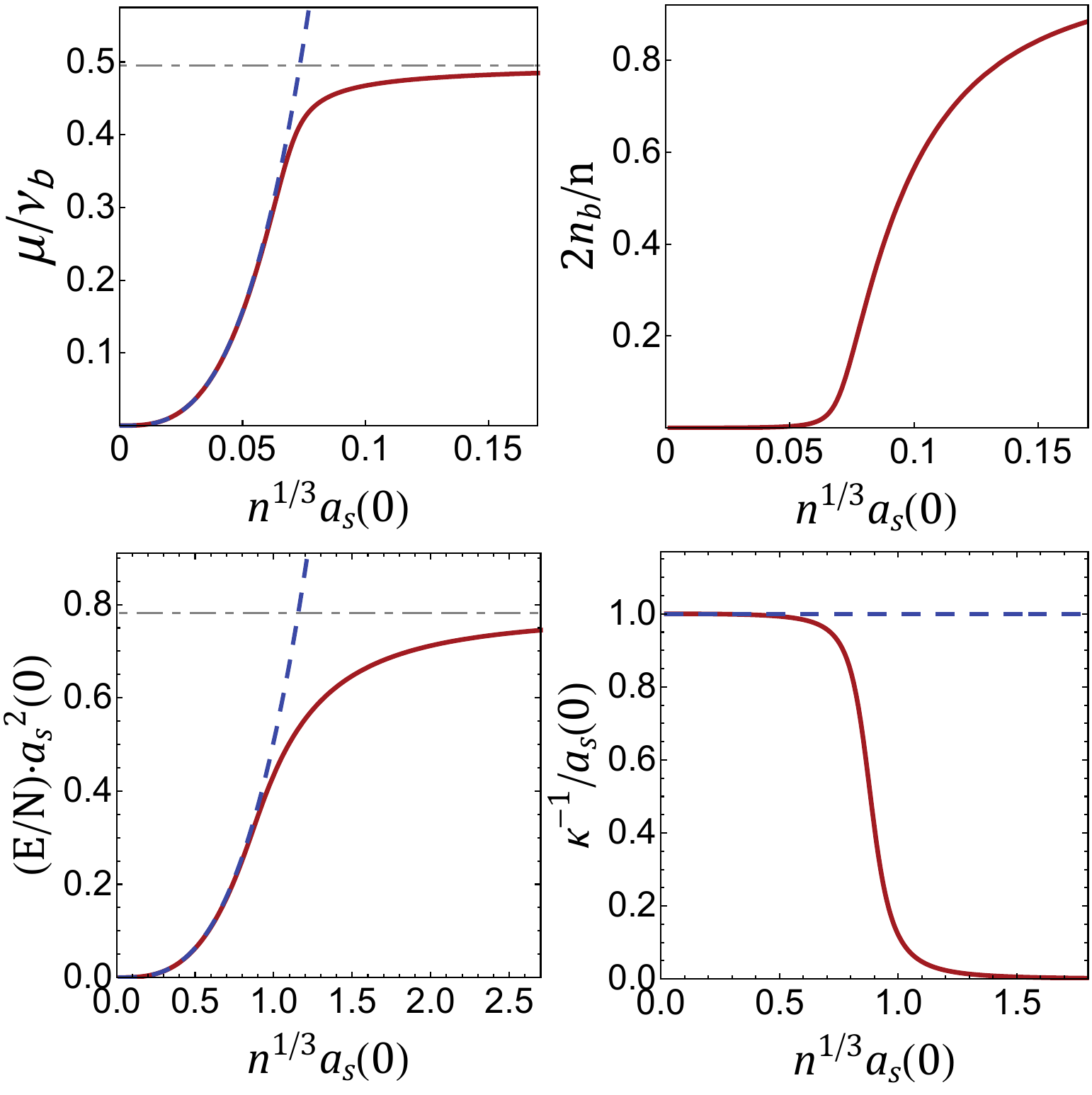}
    \put(2,98){(a)}
    \put(52,98){(b)}
    \put(2,49){(c)}
    \put(52,49){(d)}
  \end{overpic}
  \caption{$\textbf{(a)}$ The chemical potential, $\textbf{(b)}$ the proportion of molecules, $\textbf{(c)}$ the energy per particle, and $\textbf{(d)}$ the inverse compressibility $\kappa^{-1}$ as a function of total density n for narrow resonance (red lines) with $\tilde\Delta=|\alpha_r|^2/(g_r\nu_r)=0.01$ and $\nu_ra_s^2(0)=0.01$, and for broad resonance (blue deshed lines) with  a constant scattering length $a_s=a_s(0)$. The gray dot-dashed lines donates the position of $\mu_c$.}\label{Fig2}
\end{figure}

We now consider a uniform system with $V_a=V_b=0$. In this case, the solution of Eq.  (\ref{GPequation}) for the ground state are constants, and the atomic density $n_a$ and molecular density $n_b$ are given by
 \begin{align}
   n_a &=|\phi_0|^2= \frac{\mu}{g(2\mu)},\label{density uniform1}\\
   n_b &=|\beta_0|^2= \frac{1}{2}\left[\frac{1}{g(2\mu)}\frac{\mu |\alpha_r|}{2\mu-\nu_r}\right]^2,\label{density uniform2}
 \end{align}
where we defined an energy-dependent interacting strength $g(E)=4\pi\hbar^2a_s(E)/m$. It is easy to see that both $n_a$ and $n_b$, and thus the total atomic density $n=n_a+2n_b$ diverge as $g(2\mu)\rightarrow0$ corresponding to $\mu\rightarrow\mu_c=E_c/2=(\nu_r-|\alpha_r|^2/g_r)/2$. As a result, the chemical potential $\mu$ saturates to $\mu_c$ as $n$ increases. This behavior is shown exactly in Fig. \ref{Fig2}(a).

Substituting Eq. (\ref{density uniform1}) and (\ref{density uniform2}) into the mean-field action (\ref{action mf}), we obtain the total energy per particle
\begin{equation}
  \frac{E}{N} = \frac{g(2\mu)n}{2}(1-\gamma_b^2),\label{meanEperN}
\end{equation}
where $\gamma_b=2n_b/n$ is the molecular fraction representing the fraction of atoms occupying the closed channel. The behavior of $\gamma_b$ as a function of $n$ is shown in Fig. \ref{Fig2}(b). Similar to the behavior of chemical potential, the energy per particle also saturates to $\mu_c$ as the density increases as shown in Fig. \ref{Fig2}(c). These behaviors of chemical potential and energy are in qualitative difference with that given by Eq. (\ref{LHYwide}) where both quantities increase monotonically with density.

This difference can be attributed to the energy dependence of scattering length. As discussed in Sec.\ref{sec:level1}, $\mu$ and $E/N$ should roughly be determined by the effective interacting strength given by $a_{\mathrm{eff}}\simeq a_s(2\mu)$. For a broad resonance, $a_s$ is a constant and the energies increase monotonically as $n$ increases according to Eq. (\ref{LHYwide}) and (\ref{LHYwidemu}). When close to a narrow resonance, if density $n$ increases, then $\mu$ also increases such that $a_{\mathrm{eff}}$ decreases as shown in Fig. \ref{Fig1}. This decreasing of $a_{\mathrm{eff}}$ suppresses the further increasing of $\mu$ and $E/N$. Since $a_{\mathrm{eff}}$ approaches zero as $\mu\rightarrow\mu_c=E_c/2$, one may expect that $\mu$ saturates to $\mu_c$ as $n$ increases which is indeed the case as shown in Fig. \ref{Fig2}.

Now we analyze the behavior of $\mu$ and $E$ at different limits in detail. In the ultra low-density limit where we have $\mu\ll\mu_c$, the open channel dominates the scattering since the closed channel is nearly unoccupied. As a result, $\gamma_b$ tends to zero and Eq. (\ref{meanEperN}) reduces to the broad resonance result. This is shown in Fig. \ref{Fig2} where the curves for narrow resonance and broad resonance approach each other in the ultra low-density regime. As the density increases, the system reaches the opposite limit $\mu\to\mu_c$ in which the effective scattering length $a_{\mathrm{eff}}$ vanishes. This leads to a saturated energy per particle as well as a vanishingly small inverse compressibility $\kappa^{-1}$ as shown in Fig. \ref{Fig2}(d) where the compressibility $\kappa$ is given as
\begin{equation}
  \kappa = \left(\frac{\partial n}{\partial \mu}\right)_{N} = \frac{n}{\mu}\left(1+\gamma_b\frac{3\mu_c-\mu}{\mu_c-\mu}\right).\label{kappa}
\end{equation}
For a very narrow resonance with $|\alpha_r|^2/(g_r\nu_r)\ll 1$, the crossover between the above two limits roughly takes place around $n=\nu_r/g_r$.

\begin{figure}[htbp]
\centering
\includegraphics[width=1\linewidth]{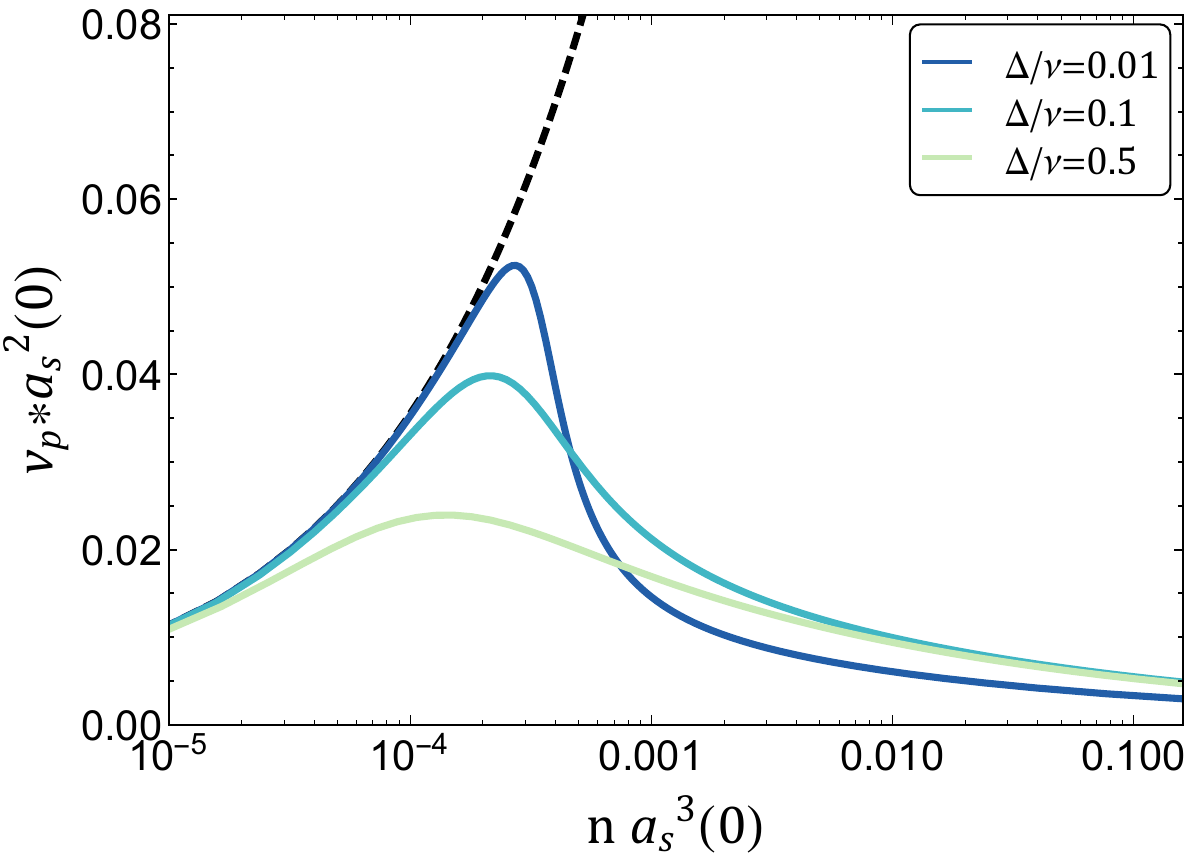}
\caption{The phonon velocity as a function of the total density $n$ for a broad resonance (black dashed line) and for narrow resonances (solid lines) with $\nu_ra_{s}^2(0)=0.01, \tilde\Delta=\{0.01, 0.1, 0.5\}$.}\label{Fig3}
\end{figure}

Furthermore, the sound velocity $v_p$ of this system can be obtained through the thermodynamic relation $v_p^2=n\kappa^{-1}$. The results are shown in Fig. \ref{Fig3}. We can see that $v_p$ develops a pronounced peak as density increases. We have verified that this value of $v_p$ is in fully consistent with the value obtained directly from the quasi-particle spectrum given in Eq. (\ref{quasiexcitation}) through
\begin{equation}
  v_p =  \lim_{k\to 0}\frac{\omega^{-}_{\mathbf{k}}}{k}.
\end{equation}

\section{\label{sec:level4}Gaussian fluctuation calculation}
\begin{figure*}[htbp]
\centering
\begin{overpic}[width=1\linewidth]{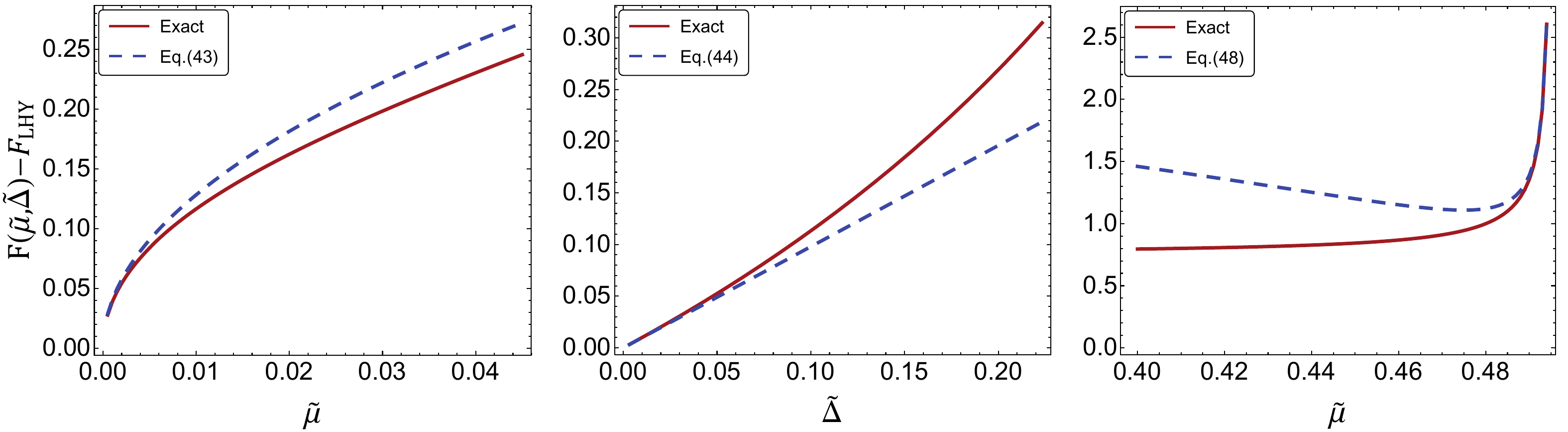}
  \put(31,6.2){(a)}
  \put(63.5,6.2){(b)}
  \put(95.5,6.2){(c)}
\end{overpic}
\caption{Behavior of $F(\tilde\mu,\tilde\Delta)-F_{\mathrm{LHY}}$ when close to three different limits. \textbf{(a)} The broad resonance limit with $\tilde\Delta=0.4$. \textbf{(b)} The single channel limit with $\tilde\mu=0.2$. \textbf{(c)} The saturation limit with $\tilde\Delta=0.01$.
The red lines in \textbf{(a-c)} show the exact results given by Eq. (\ref{Fexact}), while blue dashed lines are the asymptotic results in the corresponding limits given by Eq. (\ref{Fbroad}), Eq. (\ref{Fnarrow}) and Eq. (\ref{Fsaturation}).}\label{Fig4}
\end{figure*}

In this section, we take the correction of Gaussian fluctuation into account, and the action is governed by Eq. (\ref{action diagonalized}) which gives the thermodynamic potential
 \begin{align}
    \frac{\Omega}{V} =& \frac{\Omega_0}{V} + \frac{1}{V\beta}\sideset{}{'}\sum_{\mathbf k} \ln\left(1-e^{-\beta\omega_{\mathbf k}^-}\right)\nonumber\\
    &+ \frac{1}{V\beta}\sideset{}{'}\sum_{\mathbf k} \ln\left(1-e^{-\beta\omega_{\mathbf k}^+}\right).
  \end{align}
At zero temperature, the ground state thermodynamic potential $\Omega_0$ is given by
 \begin{align}
   \frac{\Omega_0}{V}\ =\ \frac{F_L}{V} + \frac{F_\mathrm{fluc}}{V}.\label{Fexact}
 \end{align}
 $F_L$ is the mean-field thermodynamic potential given as
 \begin{align}
   \frac{F_L}{V}\ =&\ -\frac{\mu^2}{2g(2\mu)}.
 \end{align}
 $F_\mathrm{fluc}$ is the zero-point energy correction induced by Gaussian fluctuation. After the renormalization procedure, we obtain
 \begin{align}
    \frac{F_\mathrm{fluc}}{V}
    \ =&\ \frac{1}{2V}\sideset{}{'}\sum_{\mathbf k}\bigg[
    (\omega_{\mathbf k}^-+\omega_{\mathbf k}^+)\nonumber\\
    &\left.-(\epsilon_{\mathbf k}^a+2g_r|\phi_0|^2+\epsilon_{\mathbf k}^b)+\frac{\mu^2}{2\varepsilon_{\mathbf k}}\right]\nonumber\\
    =&\ \frac{1}{4\pi^2}\left(2m\right)^{3/2}\mu^{5/2}F(\tilde\mu,\tilde\Delta).
 \end{align}
where the term $\mu^2/2\varepsilon_{\mathbf k}$ in the square brackets comes from the renormalization of the bare parameters and cancels the divergence in the momentum summation $\sum'_{\mathbf k}$.
Here $F(\tilde\mu,\tilde\Delta)$ is a dimensionless function defined as
 \begin{align}
    F(\tilde\mu,\tilde\Delta) = \int_0^\infty \left[\frac{1}{2}+x^2g(x,\tilde\mu,\tilde\Delta)\right]\mathrm dx,
 \end{align}
where $\tilde\mu = \mu/\nu_r$, $\tilde\Delta = \Delta/\nu_r$, $\Delta = |\alpha_r|^2/g_r$, and $g(x,\tilde\mu,\tilde\Delta)$ is another dimensionless function defined as
 \begin{align}
    g(x,\tilde\mu,\tilde\Delta) =& \sqrt{\omega_a^2+\omega_b^2+4\xi-1+2\sqrt{(\omega_a\omega_b-2\xi)^2-\omega_b^2}}\nonumber\\
    &-\omega_a-\omega_b,
 \end{align}
with
\begin{subequations}
  \begin{align}
    \omega_a\ &=\ x^2+1-2\frac{\tilde\Delta}{2\tilde\mu-1+\tilde\Delta},\\
    \omega_b\ &=\ \frac{x^2}{2}+\frac{1}{\tilde\mu}-2,\\
    \xi\ &=\ \frac{\tilde\Delta}{\tilde\mu}\frac{2\tilde\mu-1}{2\tilde\mu-1+\tilde\Delta}.
  \end{align}
\end{subequations}
In the broad resonance limit ($\alpha_r,~\nu_r\rightarrow\infty$ with fixed $a_s(0)$ such that $\tilde{\mu}\rightarrow 0$ with finite $\tilde\Delta$) or the single channel limit ($\alpha_r\rightarrow 0$ such that $a_s(E)= mg_r/(4\pi)\equiv a_{bg}$ and we have $\tilde\Delta\rightarrow 0$ with $\tilde{\mu}$ remaining finite), $F$ approaches a constant given by $F_{\mathrm{LHY}}=8\sqrt{2}/15$. We have verified that in these two limits, the energy per particle obtained from our Eq. (\ref{ECorr}) and (\ref{densitytotalCorr}) recovers the LHY result in Eq. (\ref{LHYwide}) (see also the comparison in Fig. \ref{Fig4}). This should be the case since in the above two limits $a_s(E)$ becomes a constant without any energy dependence.

Below, we analyze the asymptotic behavior of $F(\tilde\mu,\tilde\Delta)$ close to the above two limits as well as to the saturation limit where $\mu\rightarrow\mu_c$ as discussed in Sec. \ref{sec:level3}.

(i) In the limit $\tilde\mu\ll 1$ while $\tilde\Delta$ remains finite, which is called the broad resonance limit, we have
\begin{equation}
  F(\tilde{\mu}, \tilde{\Delta})=\frac{8 \sqrt{2}}{15}+\frac{\sqrt{6} \pi}{4} \frac{\tilde{\Delta}}{1-\tilde{\Delta}} \sqrt{\tilde{\mu}}+O(\tilde{\mu}).\label{Fbroad}
\end{equation}

(ii) In the limit $\tilde\Delta\ll 1$ which is called the single channel limit, we have
\begin{align}
  F(\tilde{\mu}, \tilde{\Delta}) &=\frac{8 \sqrt{2}}{15}+\frac{G(\tilde{\mu})}{1-2 \tilde{\mu}} \tilde{\Delta}+O\left(\tilde{\Delta}^{2}\right)\label{Fnarrow},\\
  G(\tilde{\mu}) &=\int_{0}^{\infty} \eta(x, \tilde{\mu})\mathrm{d}x,
\end{align}
where
 \begin{align}
    \eta(x,\tilde\mu) =& \frac{1}{\sqrt{(x^2+1)^2-1}+(x^2+1)}\cdot\nonumber\\
    &\frac{2x^2}{\sqrt{(x^2+1)^2-1}+\frac{x^2}{2}-2+\frac{1}{\tilde\mu}}\cdot\nonumber\\
    &\left(1+\frac{x^2}{2\sqrt{(x^2+1)^2-1}}\right).
  \end{align}
If $\tilde\mu$ also approaches 0 in this case, then we have
\begin{equation}
  G(\tilde\mu) = \frac{\sqrt 6\pi}{4}\sqrt{\tilde\mu}+O(\tilde\mu).
\end{equation}

(iii) In the limit $\tilde{\mu} \to(1-\tilde\Delta) / 2-0^+$ such that $g(2\mu)\to 0^+$ which is called the saturation limit (since this is the limit where $\mu$ and $E$ saturate to $\mu_c$), we have
\begin{equation}
  F(\tilde{\mu}, \tilde{\Delta})=\frac{\pi \sqrt{\tilde\Delta}}{4} \frac{1}{\sqrt z}+\frac{1-(21-8 \sqrt 6) \tilde\Delta}{8(1-\tilde\Delta) \sqrt{\tilde\Delta}} \pi \sqrt z+O\left(z^{3/2}\right),\label{Fsaturation}
\end{equation}
where $z=(1-\tilde{\Delta})/2-\tilde{\mu}$. The comparison between the above three asymptotic behaviors and the full results of $F$ is shown in Fig. \ref{Fig4}.

\begin{figure}[htbp]
\centering
\includegraphics[width=1\linewidth]{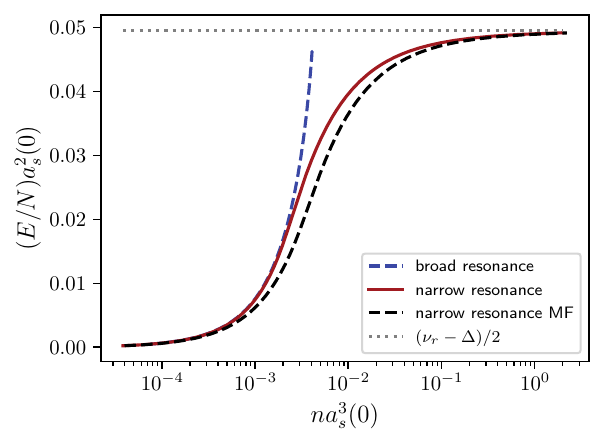}
\caption{Energy per particle as a function of density $n$ at $\nu_ra_s^2(0)=0.1,\ \Delta\cdot a_s^2(0)=0.001$. Close to a narrow resonance, $E/N$ with Gaussian fluctuation correction (red line) coincides with that of broad resonance case with a constant $a_s=a_s(0)$ (blue dashed line) at very small density and saturates to $\mu_c$ (indicated by the dotted line) at the saturation limit. The black dashed line donates the mean-field results.}\label{Fig5}
\end{figure}

The correction of energy per particle due to the Gaussian fluctuation is given by
\begin{align}
  \frac{E}{N}\ &=\  \frac{1}{n}\left[-\frac{\mu^2}{2g(2\mu)}+\frac{\left(2m\right)^{\frac{3}{2}}}{4\pi^2}\mu^{\frac{5}{2}}F(\tilde\mu,\tilde\Delta)\right]+\mu,\label{ECorr}\\
  n\ &=\ n_0 -\frac{(2\mu)^{\frac{3}{2}}}{4\pi^2}\left[\tilde\mu F'(\tilde\mu,\tilde\Delta)+\frac{5}{2}F(\tilde\mu,\tilde\Delta)\right].\label{densitytotalCorr}
\end{align}
where $n_0=n_a+2n_b$ is the mean-field density given by Eq. (\ref{density uniform1}) and (\ref{density uniform2}). The results are shown in Fig. \ref{Fig5}, and one can see that the Gaussian fluctuation correction will not change the mean-field results qualitatively. At low density, again the results approximately coincide with that of broad resonance case with a constant $a_s=a_s(0)$. As density increases towards the saturation limit, the contribution of Gaussian fluctuation is small compared with the mean-field part, and the energy per particle still saturates to $\mu_c$.

\begin{figure}[htbp]
 \centering
 \begin{overpic}[width=1\linewidth]{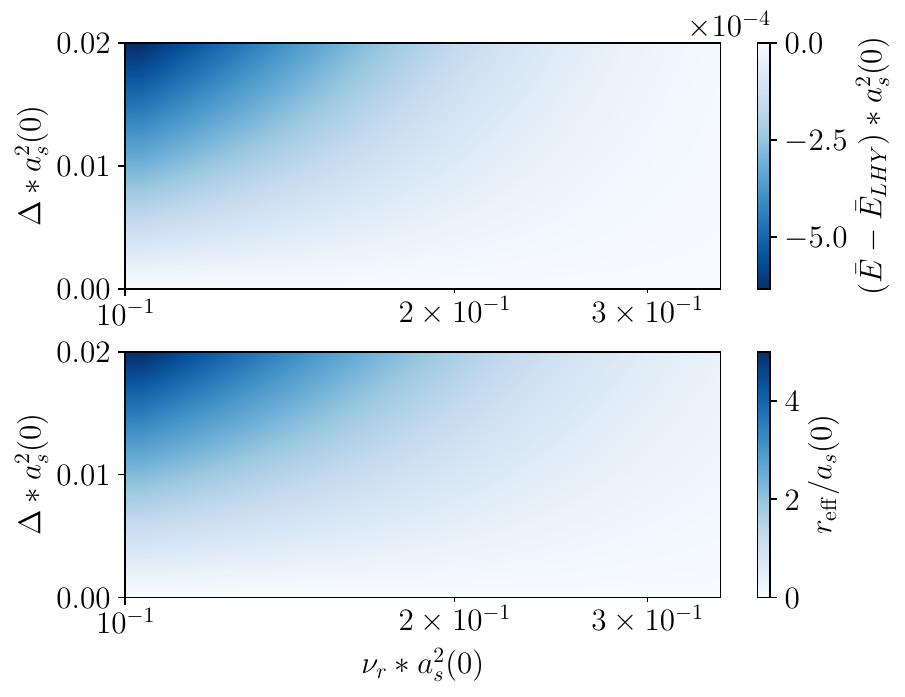}
   \put(72,48.3){(a)}
   \put(72,14){(b)}
 \end{overpic}
 \caption{$\textbf{(a)}$ Energy per particle as a function of $\nu_r$ and $\Delta$ at fixed $n^{1/3}a_s(0)=0.1$. The colorbar shows the difference with the single channel limit. $\textbf{(b)}$ $r_{\mathrm{eff}}$ as a function of $\nu_r$ and $\Delta$ at fixed $n^{1/3}a_s(0)=0.1$.}\label{Fig6}
\end{figure}

In Fig. \ref{Fig6}(a), we show $\Delta\bar{E} = E/N - E_{\mathrm{LHY}}/N$ at fixed $n^{1/3}a_s(0)$ with different detuning $\nu_r$ and resonance width $\Delta$, which is the difference between the energy per particle given by Eq. (\ref{ECorr}) and that given by LHY formula in Eq. (\ref{LHYwide}) with $a_s=a_s(0)$. Fig. \ref{Fig6}(b) shows the corresponding effective range $r_{\mathrm{eff}} = 2\Delta/\left[a_{bg}(\Delta-\nu_r)^2\right]$.
The quantity $r_{\mathrm{eff}}$ characterizes the energy dependence of scattering length, i.e. with larger $r_{\mathrm{eff}}$ the scattering length depends more sensitively on $E$. One can see that $|\Delta\bar{E}|$ vanishes in the limit $\Delta/\nu_r\ll 1$ and increases as $\Delta$ increases or $\nu_r$ decreases, which is approximately consistent with the behavior of $r_{\mathrm{eff}}$.

\begin{figure}[htbp]
\centering
\includegraphics[width=1\linewidth]{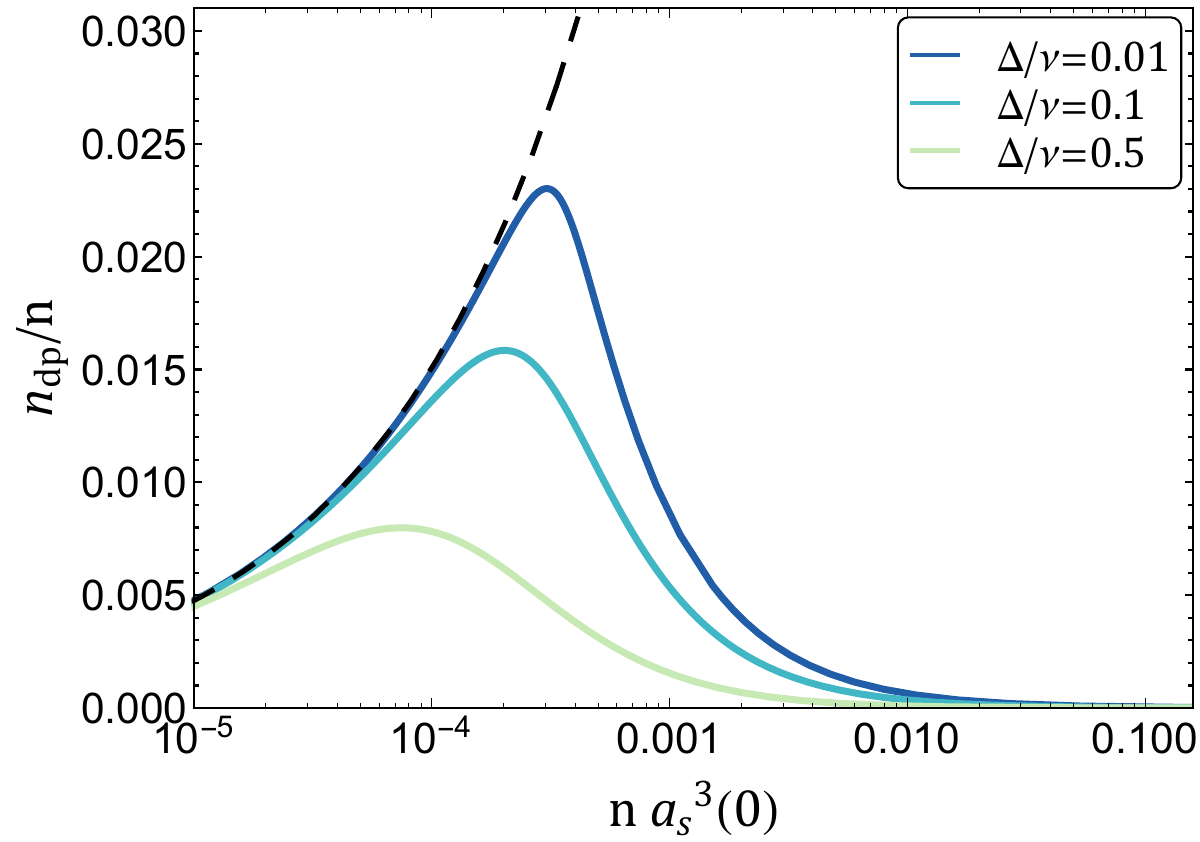}
\caption{The ratio between the density of quantum depletion and the total density $n_{\mathrm{dp}}/n$ as a function of the total density n. The solid lines from top to bottom stand for $\nu_ra_s^2(0)=0.01$ with $\tilde\Delta=\{0.01, 0.1, 0.5\}$ respectly. The black dashed line shows $n_{\mathrm{dp}}/n$ for single-channel model with $\nu_ra_s^2=0.01$.}\label{Fig7}
\end{figure}

We have also investigated the quantum depletion $n_{\mathrm{dp}}=\sideset{}{'}\sum_{\mathbf{k}}\langle G|\hat a_{\mathbf k}^\dagger\hat a_{\mathbf k}+2\hat b_{\mathbf k}^\dagger\hat b_{\mathbf k}|G\rangle$ as shown in Fig. \ref{Fig7}. In particular, we find
\begin{subequations}
  \begin{align}
    \langle G|\hat a_{\mathbf k}^\dagger\hat a_{\mathbf k}|G\rangle\ &=\ \xi_1^2+v_1^2,\\
    \langle G|\hat b_{\mathbf k}^\dagger\hat b_{\mathbf k}|G\rangle\ &=\ \xi_2^2+v_2^2,
  \end{align}
\end{subequations}
where $|G\rangle$ is the ground state of quasiparticles at zero temperature, and $\xi_1,\xi_2,v_1,v_2$ are the matrix elements of the quasi-particle transformation \cite{supple}.

At very small density, the behavior of quantum depletion also approaches that of broad resonance with a constant $a_s=a_s(0)$. In the saturation limit $\left(\mu\to\mu_c\right)$, the depletion vanishes due to the vanishing of effective interaction strength represented by $a_{\mathrm{eff}}$. The overall behavior of quantum depletion is similar to that of phonon velocity in Fig. \ref{Fig3}.

\section{\label{sec:level5}Density profile in harmonic trap}
In this section, we investigated the density distribution of a narrow Feshbach resonance Bose gas trapped in a harmonic potential with $V_b(r)/2=V_a(r)=\frac{1}{2}m\omega^2 r^2$. For trapped gas with a large particle number, the density distribution can be obtained by local density approximation by replacing $\mu$ in Eq. (\ref{density uniform1}), (\ref{density uniform2}) with local chemical potential $\mu(r)=\mu_0-V(r)$ with $\mu_0$ determined by the total particle number.

\begin{figure*}[htbp]
\centering
\begin{overpic}[width=1\linewidth]{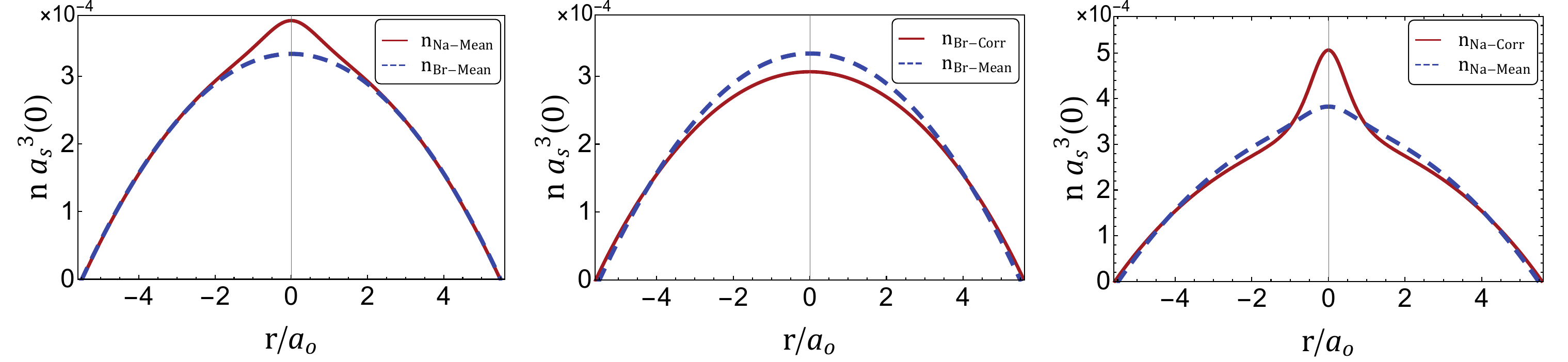}
  \put(5.7,20){(a)}
  \put(38.7,20){(b)}
  \put(71.8,20){(c)}
\end{overpic}
\caption{Density distribution $n a_s^3(0)$ of a Bose gas in a harmonic trap with fixed particle number $N=2\times 10^5$. $\textbf{(a)}$ The comparison between the narrow resonance (red line) with  $\tilde\Delta=|\alpha_r|^2/{g_r\nu_r}=0.001$ and $a_o=\sqrt{m\omega/\hbar}=60a_s(0)$ and the broad resonance (blue dashed line) with $a_s\equiv a_s(0)$ at the mean-field level. $\textbf{(b-c)}$ The comparion of the density distribution with (red lines) and without (blue dashed lines) gaussian correction for broad resonance (b) and narrow resonance (c).}\label{Fig8}
\end{figure*}

For interacting Bose gas across a broad Feshbach resonance, the distribution in a harmonic trap is approximately parabolic as a consequence of the competition between interaction energy and the trap potential \cite{TFapprox,TFapprox1}. Here, as shown in Fig. \ref{Fig8}(a), the distribution is similar to that of a broad resonance Bose gas at the edge of the trap while more particles can be accommodated in the center where the gas has higher density and thus a smaller effective scattering length $a_{\mathrm{eff}}$ as discussed in Sec. \ref{sec:level3}. A sharp peak appears if the chemical potential at the trap center $\mu_0$ approaches $\mu_c$. This density bump at the trap center is related to the vanishing of inverse compressibility in the saturation limit.

Finally, we take account of the correction by fluctuations and evaluate the density distribution by substituting $\mu$ in Eq. (\ref{densitytotalCorr}) with $\mu_0-V(r)$. The distribution compared with mean-field results with a fixed particle number is  shown in Fig. \ref{Fig8}(c). The density at the trap center is further increased after the Gaussian fluctuation is included.

\section{\label{sec:level6}Conclusion}
We studied the ground state properties of weakly interacting bose gas close to a narrow Feshbach resonance. With the help of a path integral approach, we established a low-density expansion for the equation of state of this system. As a consequence of the energy dependence for narrow Feshbach resonance, the gas behaves very differently, especially for higher density. As the density increases, the energy dependence in scattering length leads to a saturated energy, a vanishingly small inverse compressibility, and a highly suppressed quantum depletion. When the Bose gas is trapped in a harmonic potential, this effect leads to a density bump in the trap center. Such phenomena should be able to be observed in current cold atom experiments.

\begin{acknowledgments}
This project was supported by the National Key R and D Program of China Grant No. 2018YFA0306502, the NSFC under Grant No. 12022405, No. 11774426 and No. 11734010, the Beijing Natural Science Foundation (Grant No. Z180013) and by the Fundamental Research Funds for the Central Universities, and the Research Funds of Renmin University of China under Grand. No. 19XNLG12.
\end{acknowledgments}

\appendix

\section{Diagonalization of the inverse Green's function $G^{-1}$}

In this supplemental material, we provide the details on how to obtain transformation matrix $\mathbf U$ appeared in Eq. (\ref{Gmatrix}) in the main text. Considering symmetry, the transformation matrix $\mathbf U$ can be written as
\noindent
\begin{equation}
  \left(
  \begin{array}{cccc}
   \eta _2 & u_2 & v_2 & \xi _2 \\
   \eta _1 & u_1 & v_1 & \xi _1 \\
   \xi _1 & v_1 & u_1 & \eta _1 \\
   \xi _2 & v_2 & u_2 & \eta _2 \\
  \end{array}
  \right),
\end{equation}
where for convenience, we set matrix elements as real numbers. Because the quasiparticles are bosonic, it is convenient to set
\begin{equation}
  {\mathbf{U}}^T\cdot
  \begin{pmatrix}
    -1&\ &\ &\ \\
    \ &-1&\ &\ \\
    \ &\ &1&\ \\
    \ &\ &\ &1\\
  \end{pmatrix}\cdot
  \mathbf{U}=
  \begin{pmatrix}
    -1&\ &\ &\ \\
    \ &-1&\ &\ \\
    \ &\ &1&\ \\
    \ &\ &\ &1\\
  \end{pmatrix},\label{bosoniccondition}
\end{equation}
and the diagonalization of inverse Green's function is accomplished by
\begin{equation}
  {\mathbf{U}}^T\cdot
  \begin{pmatrix}
    \epsilon_b & \alpha &\ &\ \\
    \alpha & \epsilon_a & g &\ \\
    \ & g &\epsilon_a &\alpha\\
    \ &\ &\alpha &\epsilon_b\\
  \end{pmatrix}\cdot
  \mathbf{U}=
  \begin{pmatrix}
    \omega^{+}&\ &\ &\ \\
    \ &\omega^{-}&\ &\ \\
    \ &\ &\omega^{-}&\ \\
    \ &\ &\ &\omega^{+}\\
  \end{pmatrix},\label{diagonalization}
\end{equation}
where
\begin{subequations}
  \begin{align}
    \epsilon_a\ &=\ \frac{k^2}{2m}+2g\phi_0^2-\mu,\\
    \epsilon_b\ &=\ \frac{k^2}{4m}+\nu_b-2\mu,\\
    g\ &=\ g\phi_0^2+\sqrt 2\alpha\beta_0,\\
    \alpha\ &=\ \sqrt 2\alpha\phi_0.
  \end{align}
\end{subequations}
It is expedient to first take the transformation as follows,
\begin{subequations}
  \begin{align}
    u_1\ &=\ (x_1 + w_1)/2,\qquad v_1\ =\ (x_1 - w_1)/2,\\
    \eta_1\ &=\ (y_1 + z_1)/2,~\qquad \xi_1\ =\ (y_1 - z_1)/2,\\
    u_2\ &=\ (x_2 + w_2)/2,\qquad v_2\ =\ (x_2 - w_2)/2,\\
    \eta_2\ &=\ (y_2 + z_2)/2,~\qquad \xi_2\ =\ (y_2 - z_2)/2.
  \end{align}\label{substitutionxwyz}
\end{subequations}
Taking $x_2=s_1x_1,w_2=t_1w_1$ and $y_1=s_2y_2,z_1=t_2z_2$, we can obtain from Eq. (\ref{bosoniccondition}) and Eq. (\ref{diagonalization})
  \begin{subequations}
   \begin{align}
    x_1\ &=\ \left(\frac{1}{1+s_1t_1}\sqrt{\frac{\epsilon_a-g+\epsilon_b t_1^2+2\alpha t_1}{\epsilon_a+g+\epsilon_b s_1^2+2\alpha s_1}}\right)^{1/2},\\
    w_1\ &=\ \left(\frac{1}{1+s_1t_1}\sqrt{\frac{\epsilon_a+g+\epsilon_b s_1^2+2\alpha s_1}{\epsilon_a-g+\epsilon_b t_1^2+2\alpha t_1}}\right)^{1/2},
   \end{align}\label{xw}
  \end{subequations}
  \begin{subequations}
   \begin{align}
    y_2\ &=\ \left(\frac{1}{1+s_2t_2}\sqrt{\frac{\left(\epsilon-g\right)t_2^2+\epsilon_b+2\alpha t_2}{\left(\epsilon+g\right)s_2^2+\epsilon_b+2\alpha s_2}}\right)^{1/2},\\
    z_2\ &=\ \left(\frac{1}{1+s_2t_2}\sqrt{\frac{\left(\epsilon+g\right)s_2^2+\epsilon_b+2\alpha s_2}{\left(\epsilon-g\right)t_2^2+\epsilon_b+2\alpha t_2}}\right)^{1/2},
   \end{align}\label{yz}
  \end{subequations}
There are two sets of solutions of $s_1,t_1,s_2,t_2$, however, using the conditions that $u_2,v_2,\xi_2,\eta_1,\xi_1$ should tends to 0 in the limit $\alpha\to 0$, we can determine the correct solutions as follows,
\begin{equation}
  \begin{aligned}
    &s_1=-t_2 = \frac{g^2-\epsilon_a^2+\epsilon_b^2-\lambda s_0}{2 \alpha  (\epsilon_a+\epsilon_b-g)},\\
    &t_1=-s_2 = \frac{g^2-\epsilon_a^2+\epsilon_b^2-\lambda s_0}{2 \alpha  (g+\epsilon_a+\epsilon_b)},
  \end{aligned}
\end{equation}
where $\lambda$ is the sign of $g^2-\epsilon_a^2+\epsilon_b^2$ and
\begin{equation}
  s_0 = \sqrt{\left(g^2-\epsilon_a^2+\epsilon_b^2\right)^2-4 \alpha ^2 (g-\epsilon_a-\epsilon_b) (g+\epsilon_a+\epsilon_b)}.
\end{equation}
Substituting the solutions of $x,w,y,z$(\ref{xw})(\ref{yz}) into Eq. (\ref{substitutionxwyz}) we can finally obtain the transformation matrix and quasiparticle excitations.

\nocite{*}


\begin{thebibliography}{99}
\bibitem{supple}
Details on how to diagonalize the $\mathbf{G}$ matrix can be found in Appendix.
\end{thebibliography}

\end{document}